%% file: GlobularPlanet.tex
%
\input timacros.tex

\input epsf
\draft
\def\mearth{\hbox{$\hbox{M}_{\oplus}$}}
\def\msun{\hbox{$\hbox{M}_{\odot}$}}
\def\Lsun{\hbox{$\hbox{L}_{\odot}$}}
\def\gtorder{\mathrel{\raise.3ex\hbox{$>$}\mkern-14mu
             \lower0.6ex\hbox{$\sim$}}}
\def\ltorder{\mathrel{\raise.3ex\hbox{$<$}\mkern-14mu
             \lower0.6ex\hbox{$\sim$}}}

\settitle{Planets in Globular Clusters?
    }

\setauthor{Steinn Sigurdsson}
\setaffil{arXiv.org  - (originally Lick Observatory, University of California,
    Santa Cruz, CA 95064)}
\setdate{July 1992}
\setjournal{appeared in {\it Astrophysical Journal, Letters}}
\setsupport{
This research was supported in part by NASA grant NAGW--2422.
}


\setabstract{
The discovery of planets around
PSR~1257+12
suggests that planetary systems may be detected
around the recycled
pulsars found in globular clusters.
Planetary systems in dense clusters
have lifetimes to disruption
due to perturbations by
passing stars
comparable to or shorter than the pulsar
lifetime,
and observations of planets in the cores of clusters may
reveal planetary systems formally dynamically unstable on
time scales short compared to the characteristic age, $\tau_c$, of the
system.
Planets formed around cluster pulsars will most likely be restricted to
semi--major axis of $\sim 0.1-1.0\, AU$, 
while ``scavenged'' planets may be observed in wider orbits,
with no stable systems
expected in the densest clusters. Observation is most probable in
the cluster rich high--density pre--core collapse clusters such as 47Tuc.
%
}

\cover
\icover

\secbegin{Introduction}

PSR~1257+12 is a classic ``recycled'' millisecond pulsar (MSP),
with a short (6.2 ms) period, low
inferred magnetic field ($\sim 10^9\ {\rm G}$) and long characteristic age
($\tau _c \sim {\rm few}\ \times 10^8\ {\rm years}$) 
(Wolszczan 1991, Wolszczan and Frail 1992).
It is similar to the recycled pulsars observed in galactic globular
clusters (GCs) (Phinney and Kulkarni 1992, van den Heuvel 1991), and mechanisms
like those that produced the putative planets around PSR~1257+12
may generate planets
around around cluster pulsars.
Various scenarios have been proposed to account
for the presence of planets around pulsars.
In clusters
formation scenarios involving SNIIs (Lin et al. 1992)
or HMXBs (Wijers et al. 1992, Podsiadlowski et al. 1992) can not be at work,
only reaccretion from a disrupted companion or an excretion disk,
or, possibly, scavenging of
a companion's planetary system during a LMXB phase,
can apply (Tavani and Brookshaw, 1992, Stevens, et al. 1992, Krolik, 1991).
We consider the possibility of planet formation or capture
around pulsars in globular clusters, and possible parameters of
observable systems, given lifetime constraints.

\secbegin{Constraints on Formation}

Theories of planet formation require planetesimals to aggregate out
of a thin disk of dust sedimented out of the thicker accretion disk.
In the case of recycled pulsars the disk may be an ``excretion'' disk
from an ablated companion (Krolik 1991, Nakamura and Piran 1991, Tavani and Brookshaw 1992),
or the remnant of a wholly disrupted star
that merged with the neutron star (or heavy white dwarf)
(Podsiadlowski et al. 1991, Fabian and Podsiadlowski 1991). Such
disks may have considerably higher surface densities in early stages of
their evolution than the protoplanetary disks normally considered for
planet formation. The presence of PSR~1957+20 in the galaxy and 
PSR~1744-24A in the globular cluster Ter 5 (Fruchter et al., 1988, Lyne et al., 1990) suggest that this
may be a probable formation scenario.
Upon reaching a critical size, the planetesimals undergo runaway growth,
reaching masses of order $10\mearth$, before gas accretion commences,
if the ambient temperature is low enough (Cameron 1988).
The timescale for planetesimal growth is a function of metallicity,
and with cluster metallicities, $Z$, ranging from $10^{-1}$--$10^{-3}$ solar,
the presence of sufficient metals to grow planetesimals is a concern.
At $Z=10^{-3}$, the total mass of ice (rocky material) available for aggregation
is only $6\, (1.5)\, \mearth $ per $\msun $ of disk matter,
insufficient to form even a single large
planet if the total mass available in the disk is much less than $\msun$. 

The time scales for planet formation have been studied mostly in the
context of the pre--solar nebula.
Following Nakano (Nakano 1987) we estimate a conservative timescale, $t_{pf}$,
for planet formation, assuming a surface density profile,
$\Sigma (r) = f r^k$, $r \leq r_0$,
where $f$
is a normalising constant,
canonically $k=-3/2$. For total disk mass $m_d$,
$f = (k+2)m_d/2\pi r_0^{k+2}$, $k \not= -2$. Assuming
the planet formation timescale is dominated by the planetesimal
aggregation time scale, $t_{pf} \approx  10^3 {r\over {\Sigma_s (r)}}$,
where $\Sigma_s$ is the surface density of solids. $\Sigma_s (r) \approx
1.8\times 10^{-2} (4.2\times 10^{-3} )\,Z\, \Sigma (r)$ for ice and
C--Si--Fe grains respectively, for $r$ such that the local temperature is low enough for
the respective solids to condense out. Hence, for a total disk mass
$m_d = M_{-2} 10^{-2}\msun $, 
$$
t_{pf} \approx  4(16)\times 10^6 {1\over Z} {{1}\over {(k+2)M_{-2}}} r_{AU}^{1-k}\ {\rm years,}
\ceqno $$
for nucleation from ice or C--Si--Fe grains respectively,
and $r_{AU}$
is the radius in $AU$ at which planetesimal growth is taking place.
Grain formation can not take place if
the ambient temperature is greater than the vapourisation temperature of the
grains, approximately 2000K for C--Si--Fe grains, about 200K for ice grains,
providing a lower limit of $r_{min}=0.02\, (2.0) \, (L_{PSR}/\Lsun )^{1/2}\ AU$
for C--Si--Fe and ice grains respectively.
It is clear that, except possibly for the lowest metallicity clusters,
$t_{pf} \ltorder \tau_c$ for $r\gtorder r_{min}$, for C-Si--Fe nucleated planets,
assuming efficient absorption of the
pulsar flux, and pulsar luminousities characteristic of recycled pulsars.
The effects of metallicity on the surface density profile of solids
in disks are not known, it is possible that
the low opacity of metal poor accretion disks increases the timescale for
disk evolution by turbulent viscosity, leading to higher $\Sigma_{s}$ and enhanced
planet formation efficiency. 
Formation scenarios involving accretion induced collapse (Grindaly and Bailyn, 1988)
or white dwarf collisions (with another white dwarf or a neutron star),
provide cases in GCs in which the accreting matter may have a very
high metallicity and consequently a short timescale
for planetesimal aggregation.
Simulations suggest that in neutron star--binary interactions
up to half the tidal interactions may be with white
dwarfs, assuming a reasonable initial mass function and an
evolved binary population, with the heavier C--O white dwarfs
having a proportionately larger fractional cross--section for
tidal encounters (Sigurdsson and Phinney 1992).
For $L_{PSR} \sim L_{Edd}$ expected during pulsar spin--up,
$r_{min} = 6\, AU$ for C--Si--Fe nucleated planetesimals, so any planetary formation
only commences after spin--up. It is possible that any scavenged planets
present with semi--major axis $a_p \ltorder 1\, AU$ will be ablated completely during spin--up.
Most of the accretion disk may be ejected during spin--up, but
if the metals from order $10^{-2}\msun $ or greater are retained,
in an annulus from $r_{min}$ to $r_0 \ltorder 2\, AU$ enough
matter is present for
planet formation to be possible. It is not necessary to retain
the bulk of the gas, as long as the metals have opportunity to
condense out into a thin disk.
If PSR~1257+12 scavenged its planets
from a companion main--sequence star with a normal planetary system,
circularising
the orbits during the accretion phase,
a similar mechanism may also operate in clusters,
with pulsars retaining primordial
planets after disrupting the companion.
With the time scales assumed here,
primordial planetary formation
around main--sequence stars
is possible in GCs, and although the encounter between a neutron star
and a main--sequence primary may disrupt the planetary system, friction in the
envelope of the merged remnant would circularise and shrink the orbits of any surviving
planets.
This scenario permits the presence of planets in eccentric orbits with $a_p \gg 1\, AU$
around cluster pulsars in low density clusters,
whereas the timescales for reaccretion would predict
maximum $a_p \ltorder 1\, AU$ and small eccentricities.

If the cluster formation rate is dominated by neutron star--binary star
interactions (Phinney and Kulkarni 1992, Sigurdsson 1991),
and planets do not form in main--sequence binaries,
then scavenging is less likely in GCs, leaving
reaccretion from a remnant as the dominant channel for planetary formation in clusters.
Scavenging is still possible in the case where the neutron star is in a binary,
and encounters a single main--sequence star with a planetary system. If the
single star has spent a large fraction of the cluster history outside the cluster core,
it may possess an extended planetary system, even in the denser clusters. 
If pulsar recycling in GCs is dominated by binary encounters,
a companion of mass $m_c$
may remain in an
eccentric orbit, semi--major axis $a_c$,
about the accreting system (Sigurdsson 1991). The effects on 
planet formation of a stellar companion 
in a highly eccentric orbit about a protoplanetary disk
are unknown; perturbations from the companion may disrupt planet formation,
or may enhance planetary formation by providing density enhancements in the disk
and perturbing planetesimal orbits. 
Even if a stellar companion enhances planetary formation for 
wide orbit companions, close companions will prevent any
planet formation
and recycled MSPs with stellar mass companions of periods
$\ltorder 1\ {\rm year}$ would not be expected to form planets.

\secbegin{Stability of Planetary Systems in Globular Clusters}

Most cluster pulsars are found in the cores of clusters, where the stellar
density, $n_4 = n/10^4 \ {\rm pc^{-3}}$,
may be as high as $10^2$
for post--core--collapse (PCC) clusters. At such densities
close stellar encounters are frequent, and it becomes necessary to consider the
lifetime of planetary systems to disruption.
The disruption mechanism may be usefully
considered in two parts,
direct ionisation of the planet by encounters with field
stars with pericenters $p \ltorder \,a_p$,
and the perturbation of the outer planets
of a system by more distant encounters,
leading to chaotic disruption
of the planetary system through dynamical evolution of the orbits.
The cross-section for a field star of mass $M_* \sim 0.7 \msun $,
to approach to within $p_{AU}$
from a neutron star is
$\sigma \approx \pi p_{AU}^2 (1 + 36/(p_{AU}v_{10}^2))\ AU^2$,
for a relative velocity at infinity of $v_{10} = v_{\infty }/10\ {\rm km\, s^{-1}}$.
The mean time between encounters, $T$, is given by
$$
T = {{3\times 10^9}\over { < n_4 \sigma_{AU} v_{10} >}}\ {\rm years},
\ceqno $$
where $\sigma_{AU} = \sigma /\sigma (p = 1\, AU)$.
Thus a system like PSR~1257+12 would be expected to be disrupted on time scales short compared
to $\tau _c$ in PCC clusters like M15, but would not experience ionising encounters
in lower density clusters such as M13 or M53. Interestingly, the time scales for direct disruption,
in the core of the cluster,
of a planet in a $\sim 1\, AU$ orbit in a medium density cluster like 47Tuc are comparable to
$\tau_c$.
For encounters with pericenter $> a_p$, the planet's orbit is perturbed by the
encounter, with the eccentricity, $e$, and inclination being sensitive to encounters
with pericenters of order $3\, a_p$, for wider encounters the perturbation is exponentially
small and can be neglected.

The effective cross--section for ionisation and strong perturbations was evaluated
by direct integration of 25000 encounters between a
$1.4\msun \ (M_{PSR}) $ primary with a $10\mearth \ (m_p)$ secondary in a $a_p=1\, AU$,
$e=0$ orbit and a $0.7\msun \ (m_c)$ field star, with
$v_{\infty } = 5-10\ {\rm km\, s^{-1}}$. The encounters were drawn from a uniform distribution in
phase space. Defining a critical velocity $v_c$, such that for $v_{\infty } = v_c$ 
the total energy of the system is zero in the center--of--mass frame,
$$
v_c = \sqrt {G {{M_{PSR} m_p }\over {m_c }} {{(M_{PSR} + m_p + m_c )}\over {(M_{PSR} + m_p )}}
{1\over a_p} },
\ceqno $$
we find for parameters relevant for cluster encounters that $v_{\infty }/v_c \sim 16-32$.
The induced perturbations in semi--major axis, eccentricity and inclination are not sensitive to the exact mass--ratio
or relative velocity. The cross-section for direct ionisation was dominated by
encounters with pericenter $p < a_p$, with approximately one quarter of those 
encounters ionising the system. An additional quarter of such close encounters
led to an increase in the planetary semi--major axis by a factor of two or greater,
leaving a system likely to experience a second close encounter on a timescale short
compared to the timescale for the initial encounter.


If there are several planets in the system, perturbations
in $e$ or $i$ of the outer planets
greater than order $10^{-2}$ may induce instability in the planetary system,
with time scales of order $10^3 P$ or greater,
where $P$ is the period of the outermost planet (Quinlan 1992).
As Figure 1 shows, encounters with $p/a_p \ltorder 3$ are effective in
%
%
inducing perturbations of that order, with
the inclination more sensitive to distant perturbations. More effort is needed
to understand the effect of perturbations of inclination of outer planets on planetary system stability.
The transition to chaotic internal dynamics may then lead to ejections of planets or
collisions between planets on time scales short compared to $\tau _c$. Colliding planets
may remain in a (high $e$) orbit, or may disintegrate. In case of disintegration, the debris may
reaccrete into a planet in a near circular orbit on time scales short compared to $\tau _c$.

The probability of finding planets around recycled pulsars in PCC clusters
such as M15 is small. In low density clusters such as M13, planets may be found
with semi--major axis as large as $10\, AU$, with $e$ increasing with $a_p$.
A metal rich, medium density cluster, such as 47Tuc or Ter 5, offers the most interesting
prospect for planet detection, with $a_p \ltorder 1\, AU$ possible,
and $e$ increasing
with $a_p$.
Observations of planets about pulsars in 47Tuc type clusters could constrain planet
formation mechanisms and determine the presence of primordial planets in GCs.
It is possible that formally dynamically unstable planetary systems may be
observed, allowing direct comparison with numerical models.
With 10 MSPs now reported in 47Tuc (Manchester et al. 1991),
detection of planets is highly probable, {\it if}
planet formation or scavenging mechanisms operate in GCs.
Failure to detect planets around any
MSP in low or medium density GCs would strongly suggest
that the formation mechanisms
possible in GCs are not at work, and that low metallicity
inhibits planetary formation in disks around pulsars and
pre main--sequence stars.
\bigskip
I would like to thank E.S. Phinney, D. Lin,
P. Bodenheimer, P. Artymowicz and N. Murray for
helpful discussion.

\vfill\eject

\bigskip
\centerline{\bf Figure 1}
\epsfysize=288pt 
\epsfbox{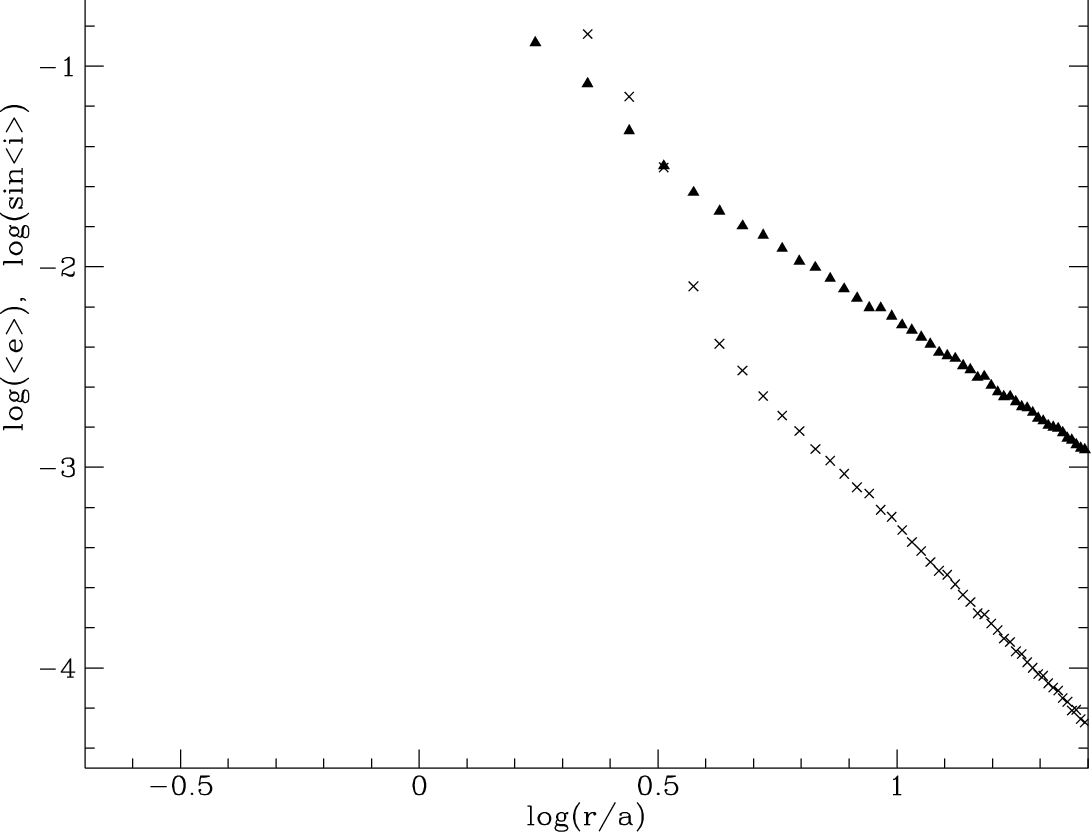}
\bigskip
\figcap{Figure 1}
Distribution of mean induced eccentricity (crosses), and inclination (triangles) relative to
original binary axis, for 25000 encounters,
as function of ratio of pericenter, $r$, to planet semi--major axis, $a$.
For $r/a$ greater than those shown perturbations in $e$ and $\sin i$ are
exponentially small.

\vfill\eject
\sectnonumber{References}

\ref\annrev{ Cameron, A.G.W., 1988}{26}{441}
\ref\Nature{Fabian, A.C. and Podsiadlowski, Ph., 1991}{353}{801}
\ref\Nature{Fruchter, A.S., Stinebring, D.R. and Taylor, J.H., 1988}{333}{237}
\ref\Nature{Grindlay, J.E. and Bailyn, C.D., 1988}{336}{48}
\ref\Nature{Krolik, J.H., 1991}{353}{829}
\ref\Nature{Lin, D.N.C., Woosley, S.E. and Bodenheimer, P.H., 1991,}{353}{827}
\ref\Nature{Lyne, A.G., Manchester, R.N., D'amico, N., Staveleysmith, L. and Johnston, S., 1990}{347}{650}
\ref\Nature{Manchester, R.N., Lyne, A.G., D'Amico, N., Bailes, M. and Lim, J., 1991}{352}{219}
\ref\Nature{Michel, F.C., 1987}{329}{310}
\ref\mn{Nakano, T., 1987}{224}{107}
\ref\apjl{Nakamura, T. and Piran, T., 1991}{382}{L81}
\ref\Nature{Phinney, E.S. and Kulkarni, S.R.}{\rm 1992}{submitted}
\ref\Nature{Podsiadlowski, Ph., Pringle, J.E. and Rees, M.J., 1991}{352}{783}
\refbook{Quinlan, G.D. 1992 in {\it Chaos, Resonance and Collective Dynamical Phenomena in the Solar System}, ed. Ferraz--Mello, S.}
\refbook{ Sigurdsson, S. 1991, {\it Ph. D. Thesis}, Caltech}
\refindent{ Sigurdsson, S. and Phinney, E.S.,}{ 1992,}{ in preparation.}
\ref\mn{Stevens, I.R., Rees, M.J. and Podsiadlowski, Ph., 1992}{254}{19p}
\ref\Nature{Tavani, M. and Brookshaw, L., 1992}{356}{320}
\ref\Nature{Thorsett, S.E. and Nice, D.J., 1991}{353}{731}
\refbook{ van den Heuvel, E.P.J. 1991, in {\it Neutron Stars: Theory and Observation}, eds. Ventura, J. and Pines, D., Dordrecht: Reidel, 99.}
\ref\Nature{Wijers, R.A.M.J., van den Heuvel, E.P.J., van Kerkwuk, M.H.
and Bhattacharya, D., 1992}{355}{593}
\ref\Nature{Wolszczan, A. and Frail, D.A., 1992}{355}{145}
\ref\Nature{Wolszczan, A., 1991}{350}{688}

\vfill\eject
\bye

%% file: timacros.tex

\input qmsfont



\def\ctr#1{\hfill#1\hfill}

\def\linebreak{\hfil\break}


\def\mnstyle{
  \def\aa{{ A \& A}}
  \def\aj{{ AJ}}
  \def\annrev{{ ARA\&A}}
  \def\apj{{ ApJ}}
  \def\apjl{{ ApJ}}
  \def\apjs{{ ApJS}}
  \def\cm{{ Celestial Mechanics}}
  \def\mn{{ MNRAS}}
  \def\Nature{{ Nature}}
  \def\science{{ Science}}
  \def\pasp{{ PASP}}
  \def\rmp{{ Rev.\ Mod.\ Phys.}}
  \def\ptms{{ Phil.\ Trans.\ Roy.\ Soc.}}
}

\def\refindent{\par\penalty-100\noindent\parskip=4pt plus1pt
               \hangindent=3pc\hangafter=1\null}
\def\ref#1#2#3#4{\refindent#2, {\it #1\/,\ }{\bf#3}, #4.}

\def\refbook#1{\refindent#1}

\def\bysame{\hbox to 50pt{\leaders\hrule height 2.4pt depth -2pt\hfill .\ }}


\def\deg{\null^{\circ}}

\def\gtorder{\mathrel{\raise.3ex\hbox{$>$}\mkern-14mu
             \lower0.6ex\hbox{$\sim$}}}

\def\ltorder{\mathrel{\raise.3ex\hbox{$<$}\mkern-14mu
             \lower0.6ex\hbox{$\sim$}}}

\def\msun{\hbox{$\hbox{M}_{\odot}$}}


\def\title#1{\vskip 24pt plus 12pt minus 12pt\tabskip0pt plus 1000pt
   \halign to \hsize{\titlefnt\ctr{##}\cr#1\crcr}\bigskip}
     
\def\author#1{\bigskip\tabskip0pt plus 1000pt \halign to \hsize{\sc\ctr{##}\cr#1
   \crcr}}

     
\def\date#1{\bigskip\centerline{\rm #1}\bigskip}
     
\def\abstract{
   \sectnonumber Abstract\par \leftskip=20pt\rightskip=20pt}




\newcount\secno         
\newcount\subsecno      
\newcount\subsubsecno
\newcount\equationno    
\def\secbegin#1{
  \vskip 0.3cm
  \goodbreak
  \noindent
  \global\advance\secno by1
  \global\subsecno=0    
  \global\equationno=0  
  \global\subsubsecno=0
  {\secfnt \the\secno.
   \hskip 2pt #1}
  \par
  \vskip 0.3cm}
\def\subsecbegin#1{
  \global\advance\subsecno by 1
  \subsubsecno=0 
  \vskip 0.3cm
  \medskip
  \goodbreak
  {\subfnt \the\secno.\the\subsecno. \hskip 2pt #1 }
  \par
  \vskip 0.3cm}
\def\subsubsecbegin#1{
  \global\advance\subsubsecno by 1
  \vskip 0.3cm
  \medskip
  \goodbreak
  {\subfnt \the\secno.\the\subsecno.\the\subsubsecno. \hskip 2pt #1 }
  \par
  \vskip 0.3cm}
\def\subsecnonum#1{
  \global\advance\subsecno by 1
  \vskip 0.3cm
  \medskip
  {\subfnt \hskip 2pt #1 }
  \par
  \vskip 0.3cm}
\def\appendix#1{
  \vskip 0.3cm
  \noindent 
  \global\equationno=0  
  {\secfnt #1 }
  \par
  \vskip 0.3cm}
\def\figcap#1{
  \smallskip
  {  \narrower\noindent #1 \smallskip } }

\def\sectnonumber#1\par{
  \global\equationno=0  
  \vskip0pt plus.2\vsize\penalty-100 
  \vskip0pt plus-.2\vsize\bigskip\vskip\parskip
  \message{#1}
  \centerline{\secfnt #1}
  \nobreak\medskip}
%

\def\ceqno{\global\advance\equationno by1
            \eqno{(\the\secno.\the\equationno)}}
\def\ceqnoa{\global\advance\equationno by1
            \eqno{(\the\secno.\the\equationno a)}}
\def\ceqnob{\eqno{(\the\secno.\the\equationno b)}}
\def\ceqnoc{\eqno{(\the\secno.\the\equationno c)}}
\def\ceqnod{\eqno{(\the\secno.\the\equationno d)}}
\def\ceqnoe{\eqno{(\the\secno.\the\equationno e)}}


\def\peqno{\global\advance\equationno by1
           \eqno{(\the\equationno)}}
\def\peqnoa{\global\advance\equationno by1
           \eqno{(\the\equationno a)}}
\def\peqnob{\eqno{(\the\equationno b)}}
\def\peqnoc{\eqno{(\the\equationno c)}}
\def\peqnod{\eqno{(\the\equationno d)}}

\def\apeqno{\global\advance\equationno by1
             \eqno{(A.\the\equationno)}}
\def\bpeqno{\global\advance\equationno by1
             \eqno{(B.\the\equationno)}}
\def\cpeqno{\global\advance\equationno by1
             \eqno{(C.\the\equationno)}}

%
\def\ceqalign{\global\advance\equationno by1
               {(\the\secno.\the\equationno)}}
\def\ceqaligna{\global\advance\equationno by1
               {(\the\secno.\the\equationno a)}}
\def\ceqalignb{{(\the\secno.\the\equationno b)}}
\def\ceqalignc{{(\the\secno.\the\equationno c)}}
\def\ceqalignd{{(\the\secno.\the\equationno d)}}
\def\ceqaligne{{(\the\secno.\the\equationno e)}}

\def\apeqalign{\global\advance\equationno by1
               {(A.\the\equationno)}}
\def\apeqaligna{\global\advance\equationno by1
               {(A.\the\equationno a)}}
\def\apeqalignb{
               {(A.\the\equationno b)}}
\def\apeqalignc{
               {(A.\the\equationno c)}}
\def\bpeqalign{\global\advance\equationno by1
               {(B.\the\equationno)}}
\def\bpeqaligna{\global\advance\equationno by1
               {(B.\the\equationno a)}}
\def\bpeqalignb{
               {(B.\the\equationno b)}}
\def\bpeqalignc{
               {(B.\the\equationno c)}}
\def\cpeqalign{\global\advance\equationno by1
               {(C.\the\equationno)}}

\newskip\ttglue
\def\draft{
   \twelvepoint
   \normalbaselineskip=24pt plus 2.4pt
   \normallineskip=3.6pt minus 1.2pt
   \normallineskiplimit=2.4pt \normalbaselines
   \rm}

\def\tenbig#1{{\hbox{$\left#1\vbox to8.5pt{}\right.\n@space$}}}
\def\eightbig#1{{\hbox{$\textfont0=\ninerm\textfont2=\ninesy
   \left#1\vbox to6.5pt{}\right.\n@space$}}}



\def\ctr#1{\hfill#1\hfill}  

\newbox\title
\newbox\author
\newbox\journal
\newbox\grpnumber
\newbox\affiliation
\newbox\abstract
\newbox\supporta
\newbox\supportb
\newbox\date
\newbox\footer

\setbox\title=\hbox{}
\setbox\author=\hbox{}
\setbox\affiliation=\hbox{}
\setbox\journal=\hbox{}
\setbox\abstract=\vbox{}
\setbox\supporta=\vbox{\hrule width\hsize
    \vskip14pt\hrule width\hsize}
\setbox\supportb=\hbox{}
\setbox\grpnumber=\hbox{}
\setbox\date=\hbox{}
\setbox\footer=\hbox{}

\def\settitle#1{\setbox\title=\vbox{
    \tabskip0pt plus 1000pt
    \baselineskip=20pt    
    \halign to \hsize{\titlefnt\ctr{##}\cr#1\crcr}
    \normalbaselines
    }}
\def\setauthor#1{\setbox\author=\vbox{
    \tabskip0pt plus 1000pt\halign to \hsize{\it\ctr{##}\cr#1\crcr}
    }}
\def\setaffil#1{\setbox\affiliation=\vbox{
    \tabskip0pt plus 1000pt\halign to \hsize{\rm\ctr{##}\cr#1\crcr}
    }}
\def\setjournal#1{\setbox\journal=\vbox{
    \tabskip0pt plus 1000pt\halign to \hsize{\rm\ctr{##}\cr#1\crcr}
    }}
\def\setdate#1{\setbox\date=\hbox{\centerline{\rm #1}}}
\def\setfooter#1{\setbox\footer=\hbox{\centerline{\rm #1}}}
\def\setgrpnumber#1{\setbox\grpnumber=\hbox{#1}}
\def\setsupport#1{
    \setbox\supportb=\vbox{
       \hrule width1in
       \vskip2pt
       \halign to \hsize{\rm\ctr{##}\cr#1\crcr}}
    \setbox\supporta=\vbox{
         \hrule width\hsize
         \vskip2pt
         \halign to \hsize{\rm\ctr{##}\cr#1\crcr}
         \vskip2pt
         \hrule width\hsize}}
\def\setabstract#1{\setbox\abstract=\vbox{
    \par{
        \vskip0pt plus.2\vsize\penalty-100 
        \vskip0pt plus-.2\vsize\bigskip\vskip\parskip
        \message{ABSTRACT}
        \centerline{\sl ABSTRACT}
        \nobreak\medskip}
    \par\leftskip=20pt\rightskip=20pt#1\par\leftskip=0pt\rightskip=0pt
    }}

\def\grp#1{\hbox to \hsize{\it Caltech\hfill GRP-{#1}}}
\def\cover{
    \pageno=0\footline={\hfil }
    \medskip
    \vskip 24pt plus 12pt minus 12pt
    \copy\title
    \bigskip\bigskip
    \copy\author
    \medskip
    \copy\affiliation
    \vfill
    \bigskip
    \copy\date
	\medskip
    \copy\footer
	\medskip
    \box\supporta
    \eject
    \footline={\hss\tenrm\folio\hss}
    \pageno=1
    }
\def\icover{
    \medskip
    \vskip 24pt plus 12pt minus 12pt
    \copy\title
    \bigskip\bigskip
    \copy\author
    \medskip
    \copy\affiliation
    \bigskip
    \copy\journal
    \copy\abstract
    \medskip
    \vfill
    \bigskip
    \copy\date
	\medskip
    \copy\footer
	\medskip
    \box\supportb
    \eject
    }

\def\deg{\ifmmode ^{\circ}
         \else $^{\circ}$\fi}
\def\pdeg{\ifmmode $\setbox0=\hbox{$^{\circ}$}\rlap{\hskip.11\wd0 .}$^{\circ}
          \else \setbox0=\hbox{$^{\circ}$}\rlap{\hskip.11\wd0 .}$^{\circ}$\fi}
\def\arcs{\ifmmode {^{\scriptscriptstyle\prime\prime}}
          \else $^{\scriptscriptstyle\prime\prime}$\fi}
\def\arcm{\ifmmode {^{\scriptscriptstyle\prime}}
          \else $^{\scriptscriptstyle\prime}$\fi}
\newdimen\sa  \newdimen\sb
\def\parcs{\sa=.07em \sb=.03em
     \ifmmode $\rlap{.}$^{\scriptscriptstyle\prime\kern -\sb\prime}$\kern -\sa$
     \else \rlap{.}$^{\scriptscriptstyle\prime\kern -\sb\prime}$\kern -\sa\fi}
\def\parcm{\sa=.08em \sb=.03em
     \ifmmode $\rlap{.}\kern\sa$^{\scriptscriptstyle\prime}$\kern-\sb$
     \else \rlap{.}\kern\sa$^{\scriptscriptstyle\prime}$\kern-\sb\fi}
%


\catcode`@=12               
\equationno=0               
\secno=0                    
\mnstyle                   
\tenpoint
